\documentclass{jltp}
\usepackage{graphicx}
\runninghead{D.S. Golubev, A.D. Zaikin, and G.
Sch\"on}{Low-temperature Dephasing \dots} 
\begin{document}

\newcommand{\gtrsim}{\, \raisebox{-2pt}{$\stackrel{>}{\sim}$}\,}

\title{On Low-temperature Dephasing by Electron-electron Interaction\thanks
{Dedicated to P.~W\"olfle  on the occasion of his 60$^{\rm th}$
birthday.}}
\author{D.S. Golubev$^{1,3}$, A.D. Zaikin$^{2,3}$, and Gerd Sch\"on$^{1,2}$}

\address{$^1$Institut f\"ur Theoretische Festk\"orperphysik,
Universit\"at Karlsruhe,\\
 76128 Karlsruhe, Germany \\
$^2$Forschungszentrum Karlsruhe, Institut f\"ur Nanotechnologie,\\
76021 Karlsruhe, Germany\\
$^3$I.E. Tamm Department of Theoretical Physics, P.N. Lebedev
Physics Institute,\\
Leninski prospect 53, 117924 Moscow, Russia}

\maketitle

\begin{center}


\end{center}

\begin{abstract}

The quantum coherence of electrons can be probed by studying weak
localization corrections to the conductivity. Interaction effects
lead to dephasing, with electron-electron interactions being the
important intrinsic mechanism.
A controversy exists whether or not the dephasing rate, as measured in
a weak localization experiment, vanishes at low temperatures.
We review the non-perturbative analysis of this question and
some of the arguments which have been raised against it. The
compact form of the presentation should make the derivation more
transparent and accessible for discussions.
We also compare with recent experiments.


\end{abstract}

\section{INTRODUCTION}

Quantum coherence and interference are basic principles
of quantum  mechanics as well as essential for the concepts
of quantum state engineering. Loss of phase coherence
results in a suppression of interference and signals a transition
from quantum to classical behavior.
Thus, it is crucial to understand how quantum
coherence is destroyed in various physical systems.

Isolated quantum degrees of freedom preserve
phase coherence. {\it Inelastic} interactions with other
-- quantum or classical -- degrees of freedom destroy coherence.
For electrons in metals at low temperatures the
dominant intrinsic interaction is electron-electron scattering.
As the temperature is lowered further, these inelastic processes
freeze out, and
the scattering rate vanishes\cite{AGD}, $1/\tau_{\rm in}\propto T^2$.
Largely because of this property it is frequently assumed that also the
electron decoherence rate $1/\tau_{\varphi}$ should vanish as the
temperature approaches zero.

The electron decoherence can be probed by studying the weak
localization correction to the conductivity.
While many of the earlier experiments\cite{Bergmann} appeared to confirm the
theoretical expectations\cite{AAK,CS,SAI}, more recent work\cite{Webb,Mohanty}
(see Ref.\ \onlinecite{Mohanty} for further references)
suggested that the electron dephasing time $\tau_{\varphi}$ in disordered
metals saturates at sufficiently low temperatures.
Extensive experimental checks\cite{Webb} allowed to rule out various
sources for this saturation, such as electron scattering from magnetic
impurities, heating and external high frequency noise.
Several other saturation mechanisms had been explored, such as $1/f$
noise, interaction of  electrons with nuclear magnetic moments or with
two-level systems in the two-channel Kondo regime\cite{Zawadowski}.
They appear not to play a role in the experiments described in Ref.\
\onlinecite{Mohanty}, nor in the more recent
 studies of Ref.\ \onlinecite{Nat}. As concluded in Ref.\
\onlinecite{Mohanty}, these recent experimental findings
``reinforce the earlier conclusion that the saturation is a real effect,
most likely arising from electron-electron interaction''.

A low-temperature saturation of $\tau_{\varphi}$
due to electron-electron interactions
has drastic physical consequences\cite{Webb,GZL}.
One of them is that the Coulomb interaction precludes (strong) Anderson
localization in disordered conductors. Another is that the
Fermi liquid theory may not apply to disordered conductors. Dephasing
at low temperatures also imposes practical limitations
on quantum state engineering with electronic devices. All that makes
the low-temperature electron decoherence one
of the most intriguing issues of modern condensed matter
physics.

In this context it is important to note that several model
systems where a quantum 
particle is coupled to a bath (e.g.\ a bath of harmonic oscillators as
in the Caldeira-Leggett model\cite{CL})
yield results which indicate a nonvanishing dephasing rate even at 
zero temperature\cite{Weiss,Loss,GZ3,Buttiker}.
We further like to mention the work of A.G. Aronov and P.
W\"olfle\cite{Peter} which emphasizes the fundamental importance of  
dephasing due to fluctuating magnetic field for such issues
as high temperature superconductivity and the metal-insulator phase
transition.

The aim of the present paper is to review the theoretical
analysis of the Coulomb-interaction-induced quantum dephasing
of electrons in disordered conductors derived before by two of
the present authors (GZ)\cite{GZL,GZB,GZB2}. This work has stimulated
a heated discussion and 
exchange of lengthy publications. One of our goals is to present 
in a compact form the main steps of the derivation,
which should help an interested reader to judge the validity of this approach. 
Some of the more recent work  -- in particular we like to mention
that of von Delft\cite{Jan} -- have put the discussion on a more
rational basis by explicitly pointing out approximations which had
been used. The second goal of this article is to justify
these approximations. Finally, we will compare with some
recent experiments with the aim to judge whether they can help to
discriminate between different theoretical views.

The structure of the paper is as follows: We will first review the
derivation of weak localization effects on the conductivity as well
as a simple analysis of the influence of electron-electron interactions.
At this level we reproduce some important results, but,
in addition, we introduce the formalism and notations which
will be used in later sections, where the same questions will be analyzed
on a more rigorous level. We then summarize in compact form
the derivation which has been presented before
by two of us (GZ)\cite{GZL,GZB,GZB2}. This part should make
explicit which approximations have been used.
We then will present several arguments which had been raised against
the conclusions of GZ as well as our counterarguments. The final
section is devoted to a comparison with recent experiments.

\section{WEAK LOCALIZATION BASICS}

In this section we shortly review the derivation of
the weak localization correction to the conductivity for
non-interacting electrons using a path integral formulation, and we
present a hand-waving approach to
account for the effects of electron-electron interactions.

The density matrix of a quantum particle evolves according to
\begin{equation}
\rho(t_{\rm f},\mathbf{r}_{1 {\rm f}},\mathbf{r}_{2 {\rm f}})
=\int d\mathbf{r}_{1 {\rm i}} d\mathbf{r}_{2 {\rm i}} \,
J(t_{\rm f},t_{\rm i},\mathbf{r}_{1 {\rm f}},\mathbf{r}_{2 {\rm
f}},\mathbf{r}_{1 {\rm i}},\mathbf{r}_{2 {\rm i}}) \,
\rho(t_{\rm i},\mathbf{r}_{1 {\rm i}},\mathbf{r}_{2 {\rm i}}),
\label{rho0}
\end{equation}
where the time evolution operator $J$ combines the forward and
backward propagators typical for the density matrix.
For a single particle which moves in an impurity
potential $U(\mathbf{r})$ and which is
subject to a static applied electric potential $V_x$, this operator
can be expressed as a path integral over two paths
\begin{eqnarray}
J(t_{\rm f},t_{\rm i},\mathbf{r}_{1 {\rm f}},\mathbf{r}_{2 {\rm f}},\mathbf{r}_{1 {\rm i}},\mathbf{r}_{2 {\rm i}}) =
\hspace{7.5cm}
\nonumber\\
=\int\limits_{\mathbf{r}_1(t_{\rm i})=\mathbf{r}_{1 {\rm i}}}^{\mathbf{r}_1(t_{\rm f})=\mathbf{r}_{1 {\rm f}}}{\cal D}\mathbf{r}_1
\int\limits_{\mathbf{r}_2(t_{\rm i})=\mathbf{r}_{2 {\rm i}}}^{\mathbf{r}_2(t_{\rm f})=\mathbf{r}_{2 {\rm f}}}{\cal D}\mathbf{r}_2
\; {\rm e}^{iS_0[\mathbf{r}_1]-iS_0[\mathbf{r}_2]+i\int_{t_{\rm i}}^{t_{\rm f}} dt'
[eV_x(\mathbf{r}_1(t'))-eV_x(\mathbf{r}_2(t'))]} \; ,
\label{J0}
\end{eqnarray}
where
$S_0[\mathbf{r}]=\int_{t_{\rm i}}^{t_{\rm f}}dt'\,[m\dot\mathbf{r}^2(t')/2-U(\mathbf{r}(t'))]$.

We are interested in the conductivity in spatially homogeneous and
stationary situations. We, therefore, assume $V_x(\mathbf{r})=-\mathbf{Er}$
and calculate the current at $\mathbf{r}$ and $t_{\rm f}$ due to the field at
 $\mathbf{r}(t')$ and $t'$
in an expansion of the density matrix to linear order in $V_x$. Thus
we get
\begin{eqnarray}
\sigma = \frac{e^2}{3m}
\int\limits_{t_{\rm i}}^{t_{\rm f}}dt'\int
d\mathbf{r}_{1 {\rm i}}d\mathbf{r}_{2 {\rm i}}
[\nabla_{\mathbf{r}_{1 {\rm f}}}-\nabla_{\mathbf{r}_{2 {\rm f}}}]
\int\limits_{\mathbf{r}_{1 {\rm i}}}^{\mathbf{r}_{1 {\rm f}}}{\cal D}\mathbf{r}_1
\int\limits_{\mathbf{r}_{2 {\rm i}}}^{\mathbf{r}_{2 {\rm f}}}{\cal D}\mathbf{r}_2
\; [\mathbf{r}_1(t')-\mathbf{r}_2(t')]
\nonumber\\
\times\,
 {\rm e}^{iS_0[\mathbf{r}_1]-iS_0[\mathbf{r}_2]} \,
\rho(t_{\rm i},\mathbf{r}_{1 {\rm i}},\mathbf{r}_{2 {\rm i}})
\bigg|_{\mathbf{r}_{1 {\rm f}}=\mathbf{r}_{2 {\rm f}}=\mathbf{r}}\; .
\label{sigma1}
\end{eqnarray}
In the stationary limit the result does not depend on
time. Accordingly we can suppress the dependence of $\sigma$ on
the time $t_{\rm f}$ and set $t_{\rm i}\to-\infty$.
We also assume the current to be averaged over the final position
$\mathbf{r}_{1 {\rm f}}=\mathbf{r}_{2 {\rm f}}=\mathbf{r}$.
Hence, the conductivity (\ref{sigma1}) does not depend on this coordinate.

When the initial density matrix is chosen to be the
equilibrium form $\rho_0$  of non-interacting electrons
-- which does not evolve in time -- the
expression for the conductivity can be simplified further.
Since Eq. (\ref{sigma1}) contains no time dependence
for times earlier than $t'$, the path integral can be
restricted to times later than $t'$. Hence we get
\begin{eqnarray}
\sigma = \frac{e^2}{3m}\int\limits_{-\infty}^{t_{\rm f}}dt'\int
d\mathbf{r}'_{1}d\mathbf{r}'_{2} &
[\nabla_{r_{1 {\rm f}}}-\nabla_{r_{2 {\rm f}}}] \;
J_0(t_{\rm f},t',\mathbf{r}_{1 {\rm f}},\mathbf{r}_{2 {\rm f}},\mathbf{r}'_{1},\mathbf{r'}_{2})
\nonumber\\
&\times\,
[\mathbf{r}'_{1}-\mathbf{r}'_{2}] \;
\rho_0(\mathbf{r}'_{1},\mathbf{r}'_{2})
\bigg|_{\mathbf{r}_{1 {\rm f}}=\mathbf{r}_{2 {\rm f}}} \;,
\label{sigma2}
\end{eqnarray}
where the time evolution operator $J_0$ is obtained from
Eq.\ (\ref{J0}) by setting $V_{\rm x}=0.$
The expression (\ref{sigma2}) can be shown to be equivalent to
the standard Kubo formula for non-interacting electrons
(see e.g.\ Ref.\ \onlinecite{Imry}).

In metals, where the electron wavelength is short, one can evaluate
the path integral $J_0$ in the `quasiclassical' approximation. This
assumes that the main contributions arise from classical paths
$\mathbf{r}^{\rm cl}(t)$ and quadratic
fluctuations around them. In the present problem we have to specify the
paths on both branches of the `Keldysh contour', $\mathbf{r}_1^{\rm
cl}(t)$ and $\mathbf{r}_2^{\rm cl}(t)$. The main
contributions to the conductivity (\ref{sigma2}) arise from
pairs of identical classical paths, $\mathbf{r}_1^{\rm
cl}(t)=\mathbf{r}_2^{\rm cl}(t)$. They yield the Drude conductivity
$ \sigma_{\rm D}= 2e^2N_0 D$, where
$D$ is the diffusion constant, proportional to the elastic
mean free time $\tau_{\rm e}$.  As shown by Altshuler, Aronov and
Khmelnitsky (AAK)\cite{AAK} and emphasized by Chakravarty and
Schmid (CS)\cite{CS}, the first quantum correction to the 
conductivity, due to `weak localization', is derived from
pairs of time-reversed classical paths $\mathbf{r}_2^{\rm
cl}(t'+t)=\mathbf{r}_1^{\rm cl}(t_{\rm f}-t)$.
They yield the correction
\begin{equation}
\delta\sigma_{\rm WL}^{0}=-\frac{2e^2D}{\pi}
\int\limits_{\tau_{\rm e}}^{\infty} dt\, W(t),
\label{sigma21}
\end{equation}
where $W(t)=(4\pi Dt)^{-d/2}a^{d-3}$ is the classical
probability for a diffusive particle to return to an initial point
after time $t$. Here $d$ is the dimensionality of the sample and
$a$  the film thickness for $d=2$, or the square root of the wire
cross section (area) for $d=1$.
The weak localization correction diverges in one and two dimensions
due to the contributions from the upper limit of the integration.
This divergence is removed, when interactions are taken into account.
Their effect is to cut the integral at times exceeding
the `dephasing time' $\tau_\varphi$. This time should be determined in
the following.

Under ideal situations at low temperatures the dephasing time $\tau_\varphi$
is limited by electron-electron interactions. In order to
determine it we proceed in two stages. We first present
a hand-waving approach in the spirit of AAK\cite{AAK} and CS\cite{CS},
which is accepted to be sufficient at
not too low temperatures. In the next section we will substantiate
and improve the derivation.

The essence of the simple approach is to replace
the interaction by an equivalent Nyquist noise source. I.e.\
the electrons move in the effective
fluctuating Gaussian potential $\delta V(t,\mathbf{r})$ with the
correlator
\begin{eqnarray}
\langle \delta V(t,\mathbf{r})\delta V(0,0)\rangle
\equiv I(t,\mathbf{r})
=\int\frac{d\omega d^3k}{(2\pi)^4}\,
{\rm Im}\left(\frac{-4\pi \coth(\omega/2T)}{k^2\epsilon(\omega,k)} \right)\,
{\rm e}^{-i\omega t+i\mathbf{k}\mathbf{r}},
\label{noise}
\end{eqnarray}
which is fixed by the fluctuation-dissipation theorem.
Here $\epsilon(\omega,k)$
is the dielectric susceptibility of the material.
The fluctuating potential should be added\cite{CS} to the action
of the electron,
$
S_0\to S=\int_{t_{\rm i}}^{t_{\rm f}}dt\,[m\dot\mathbf{r}^2/2-U(\mathbf{r})
+e\delta V(t,\mathbf{r})], $
which enters, e.g., the expression for the conductivity (\ref{sigma1}).

In the presence of the Nyquist noise a reduction of the time span of
the path integration, as done in the transition from
(\ref{sigma1}) to (\ref{sigma2}), is no longer rigorously allowed since
the density matrix now depends on time.
Any arbitrary density matrix at  $t_{\rm i} = - \infty$ should
relax to the equilibrium density matrix of the interacting problem.
The use of a reduced form equivalent to (\ref{sigma2})
with, e.g., the equilibrium density matrix of the non-interacting problem
at time $t'$, in general may lead to spurious results describing
the decay of this specific initial state. However, as will be shown below,
for the weak localization correction
this approximation is justified with sufficient accuracy.

Due to the Gaussian nature of the fluctuating field
$\delta V$ one can perform the average and finds that the exponent in
Eq.~(\ref{sigma1}) acquires an imaginary part, 
\begin{equation}
{\rm e}^{iS_0[\mathbf{r}_1]-iS_0[\mathbf{r}_2]
-S_{\rm I}[\mathbf{r}_1,\mathbf{r}_2]}\; ,
\label{sigma3}
\end{equation}
with
\begin{eqnarray}
S_{\rm I}=\frac{e^2}{2}\int\limits_{t_{\rm i}}^{t_{\rm f}}dt\int\limits_{t_{\rm i}}^{t_{\rm f}} d\tilde t
\;\big[ I(t-\tilde t,\mathbf{r}_1(t)-\mathbf{r}_1(\tilde t))
+ I(t-\tilde t,\mathbf{r}_2(t)-\mathbf{r}_2(\tilde t)) \nonumber\\ -\,
 I(t-\tilde t,\mathbf{r}_1(t)-\mathbf{r}_2(\tilde t))
- I(t-\tilde t,\mathbf{r}_2(t)-\mathbf{r}_1(\tilde t))\big] \; .
 \label{SI}
\end{eqnarray}

\begin{figure}
\begin{center}
\includegraphics[width=10cm]{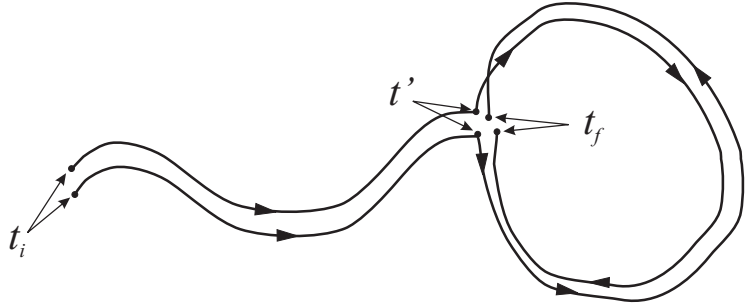}
\end{center}
\caption{A pair of classical paths which contributes to
the weak localization correction.}
\end{figure}

The effects of interactions, which are responsible for the decay of
correlations and coherence, are lumped in the imaginary part of the
action $S_{\rm I}$.
As long as the interactions are weak we can analyze their effect
within the quasiclassical approximation. For this
we need to evaluate $S_{\rm I}$ on the important
classical paths $\mathbf{r}_2(t)$ and $\mathbf{r}_1(t)$,
satisfying the classical equation of motion without interaction
$\delta S_0[\mathbf{r}]/\delta\mathbf{r}=0$.
Since we are interested in weak localization corrections,
we concentrate on paths involving pairwise time-reversed
paths. The leading contribution arises from pairs of paths
as shown in Fig.\ 1.
The two paths are equal, $\mathbf{r}_2(t)=\mathbf{r}_1(t)$, for times
earlier than $t'$, they are time-reversed,
$\mathbf{r}_2(t)=\mathbf{r}_1(t_{\rm f}+t'-t)$, for times $t'<t<t_{\rm f}$, and
they pass the same region within a diameter of order of the mean free
path $l$ at two times, $t=t'$ and $t=t_{\rm f}$.
In principle, there could be additional loops with time-reversed parts
before $t'$. However, we can ignore such contributions because
they constitute higher-order weak localization corrections
to the conductivity, which decay faster in time.
For the paths considered, it is easy to see that the double time
integral in Eq.\  (\ref{SI}) reduces to
\begin{equation}
\int_{t_{\rm i}}^{t_{\rm f}} dt \int_{t_{\rm i}}^{t_{\rm f}} d\tilde t \to
\int_{t'}^{t_{\rm f}} dt \int_{t'}^{t_{\rm f}} d\tilde t \; .
\label{replacement}
\end{equation}
The imaginary part of the action $S_{\rm I}(t_{\rm f}-t')$
grows with the time difference  $t_{\rm f}-t'$. For a given pair
of paths it may fluctuate on a short time scale of order $\tau_{\rm e}$,
however this averages out when we sum over all possible paths.

The classical paths of electrons in disordered metals are diffusive.
When we add the contributions of all possible pairs of 
time-reversed paths a decaying factor
$\langle W(t_{\rm f}-t') {\rm e}^{-S_{\rm I}(t_{\rm f}-t')}\rangle$
appears under the integral (\ref{sigma2}). Our aim is to find
the characteristic time scale $\tau_\varphi$ characterizing this decay.
In order to do that, we may replace
$\langle W(t) {\rm e}^{-S_{\rm I}(t)}\rangle$
by $W(t){\rm e}^{-\langle S_{\rm I}(t)\rangle}$.
This simple averaging has been also performed by CS\cite{CS},
while AAK\cite{AAK} evaluated directly $\langle W(t) {\rm e}^{-S_{\rm
I}(t)}\rangle$. 
However, both approaches gave the same expression for $\tau_\varphi$
up to a prefactor of order one, which is of little importance.

After these steps the weak localization correction to the conductivity
 (\ref{sigma21}) is modified to become
\begin{equation}
\delta\sigma_{\rm WL}=
-\frac{2e^2D}{\pi}\int\limits_{\tau_{\rm e}}^{+\infty}dt\,
W(t)\,{\rm e}^{-f_d(t)} \; ,
\label{sigmaWL}
\end{equation}
where $f_d(t)=\langle S_{\rm I}(t)\rangle$ has the form
\begin{equation}
f_d(t)=e^2\int\limits_0^{t} dt' \int\limits_0^{t} d t''
\int d^d x D(|t'- t''|,x) [I(t'-t'',x)-I(t'+t''-t,x)] \; .
\end{equation}
It depends on the `diffuson'
$D(t,r)=(4\pi Dt)^{-d/2} a^{d-3} \exp(-r^2/2t)$,
i.e.\ the probability that an
electron propagates  the distance $r$ within a time  $t$.

Using the Drude formula for the dielectric susceptibility of
disordered metal, $\epsilon(\omega,k)=4\pi/(-i\omega +Dk^2)$,
we can express the function $f_d(t)$ as\cite{GZB2}
\begin{eqnarray}
f_d(t)&=&\frac{\pi^{\frac{3d-d^2-2}{2}}e^2(4D)^{1-\frac{d}{2}}}
{2^{3d-d^2-2}\sigma_{\rm D}a^{3-d}}
\int\frac{d\omega\,d\omega'}{(2\pi)^2}\,
|\omega'|^{\frac{d}{2}-2}\bigg[(\omega-\omega')
\coth\frac{\omega-\omega'}{2T}
\nonumber\\
&&\times\, \frac{1-\cos\omega t}{\omega^2}
-\omega\coth\frac{\omega}{2T}\frac{\cos\omega t-\cos\omega' t}
{\omega'^2-\omega^2} \bigg] \; .
\label{fd}
\end{eqnarray}
It is obvious that $f_d(t)$ grows with time and limits the range
of time integration.
The dephasing time can be defined by the condition $f_d(\tau_\varphi)\sim 1$,
with the result\cite{GZL}
\begin{equation}
\frac{1}{\tau_\varphi}=\frac{4e^2D}{\sigma_{\rm D}a^{3-d}}
\int^{1/\tau_{\rm e}}_{1/\tau_{\varphi}}\frac{d\omega}{2\pi}
\int \frac{d^dq}{(2\pi)^d}
\frac{\omega\coth(\omega/2T)}{\omega^2+(Dq^2)^2} \; .
\label{GZ}
\end{equation}
Explicitly we find\cite{GZL,GZB}
\begin{eqnarray}
{1}/{\tau_\varphi}&=\frac{e^2}{\pi\sigma_{\rm D} a^2}
\sqrt{\frac{2D}{\tau_e}}\,
[1+2T\sqrt{\tau_\varphi\tau_e}]\; \; & {\rm in}\; {\rm 1D},
\nonumber\\
{1}/{\tau_\varphi}&=\frac{e^2}{4\pi\sigma_{\rm D} a\tau_e}
\, [1+2T\tau_e\ln(T\tau_e)]
\;\; & {\rm in}\; {\rm 2D},
\nonumber\\
{1}/{\tau_\varphi}&=\frac{e^2}{3\pi^2\sigma_{\rm D}\sqrt{2D}\tau_e^{3/2}} \,
[1+6(T\tau_e)^{3/2}]\;\; & {\rm in}\; {\rm 3D}.
\label{tauphi}
\end{eqnarray}
The divergence of the integral in (\ref{GZ}) at high frequencies
is cut at $\omega\sim 1/\tau_{\rm e}$ because a
classical path needs a time exceeding $\tau_e$ to return to the same point.
Because of the quantitative ambiguity related to this cut-off
procedure, as well as the
definition of $1/\tau_\varphi$, the results are
valid only up to numerical prefactors of order one.

We observe that the dephasing rate (\ref{GZ}) does not vanish even at
zero temperature, and, in this respect, differs from the earlier
expectations. The main difference between our work and that of AAK\cite{AAK} and
CS\cite{CS} lies in the fact that they ignore fluctuations of the potential with
frequencies higher than $T$. This procedure yields a dephasing rate
which vanishes at $T=0$. For a comparison of the results it is useful to split
  (\ref{GZ}) as
\begin{equation}
\frac{1}{\tau_{\varphi}(T)}=\frac{1}{\tau_{\varphi 0}}+\frac{1}{\tau
(T)} \,.
\label{sum}
\end{equation}
Here $\tau_{\varphi 0}$ is our -- controversial -- zero
temperature dephasing time, and $\tau (T)$ is basically the result
derived by AAK\cite{AAK}. The latter has also been supported by
Aleiner, Altshuler and Gershenson (AAG)\cite{AAG}. Within the framework of
first order perturbation theory in the interaction they arrive at
\begin{eqnarray}
\frac{1}{\tau_\varphi}=\frac{e^2D}{T\sigma_{\rm D}a^{3-d}}
\int\limits_{1/\tau_\varphi}^{\infty}\frac{d\omega}{2\pi}
\int \frac{ d^dq}{(2\pi)^{d}}\,d\xi\,
\frac{\omega[\coth({\omega}/{2T})+\tanh({(\xi-\omega)}/{2T})]}
{(\omega^2+(Dq^2)^2)\cosh^2({\xi}/{2T})}.
\label{AAG}
\end{eqnarray}
In this form, after performing the integration over $\xi$ one observes that
high frequencies $\omega>T$ do not contribute to the integral, and 
(\ref{AAG})
effectively reduces to the result obtained by AAK\cite{AAK}.
Below we will comment on the arguments for either of the results.
Before doing so we present, in the following section, a more
thorough analysis of interaction effects.

\section{INTERACTION EFFECTS: MORE RIGOROUS APPROACH}

In this Section we review a more thorough description
of interaction effects in disordered conductors, based on the work
of Refs.~\onlinecite{GZB} and \onlinecite{GZB2}, which takes into
account Pauli principle etc.
The formulation of the problem of electron-electron interaction in
metals is similar in many respects to the problem of a particle
interacting with a bath of harmonic oscillators, which is known as the
Caldeira-Leggett (CL) model.

First, the fluctuating electro-magnetic field needs to be
quantized. This is achieved by introducing two independent
fluctuating potentials $V_1$ and $V_2$ for the two branches of the
Keldysh contour (see, e.g., Ref.~\onlinecite{AES}) or, equivalently,
the combinations
$V^+=(V_1+V_2)/2$ and $V^-=V_1-V_2$. We assume the fluctuations to
be small, and expand the action up to the second order in $V^\pm.$
This approximation is equivalent to RPA. The first order term in
the fields describes the motion of electrons in a effective static
potential of other electrons. It leads to the so-called Hartree
diagrams in perturbation theory with closed electronic loops. In a
homogeneous Jellium model these terms vanish because of charge
neutrality, but in real disordered metals this is not the case.
Nevertheless, these terms are not important for dephasing, and we
ignore them, leaving in the action only the terms quadratic in
$V^\pm$. Then the properties of the electro-magnetic field are
fully determined by the correlators of $V^\pm$, i.e.\ the photon Green
functions. In homogeneous cases they have the form:
\begin{equation}
\langle V^+(t,\mathbf{r})V^+(0,0)\rangle =
I(t,\mathbf{r}), \;
\langle V^+(t,\mathbf{r})V^-(0,0)\rangle=
iR(t,\mathbf{r}), \;
\langle V^-V^-\rangle \equiv 0 \, ,
\label{VV}
\end{equation}
where
\begin{equation}
R(t,\mathbf{r})
=\int\frac{d\omega\, d^3k}{(2\pi)^4}\,\frac{4\pi}{k^2\epsilon(\omega,k)}
\,{\rm e}^{-i\omega t+i\mathbf{k}\mathbf{r}} \, ,
\label{IR}
\end{equation}
and the function $I(t,\mathbf{r})$ is defined by Eq. (\ref{noise}).
Note that the function $R(t,\mathbf{r})$ vanishes identically
for $t<0$ due to causality.

The density matrix of an interacting electron in a metal,
$\rho(\mathbf{r}_1,\mathbf{r}_2)=\langle
\Psi^+(\mathbf{r}_2)\Psi(\mathbf{r}_1)\rangle$,
can be expressed in terms of the density matrix
$\rho_V(\mathbf{r}_1,\mathbf{r}_2)$ of an electron in
the fluctuating potentials $V^\pm$ as follows:
\begin{equation}
\rho(\mathbf{r}_1,\mathbf{r}_2)
= \langle\rho_V(\mathbf{r}_1,\mathbf{r}_2) \rangle_{V^\pm} \; .
\end{equation}
The averaging over $V^\pm$ is performed with the aid of
(\ref{VV}). The matrix $\rho_V(\mathbf{r}_1,\mathbf{r}_2)$
satisfies the generalized nonlinear Liouville  equation:
\begin{equation}
i\frac{\partial\rho_V}{\partial t}
=[H_0-eV^+,\rho_V]-\frac{1}{2}\, (1-\rho_V)\, eV^-\rho_V
-\frac{1}{2}\, \rho_V \, eV^-(1-\rho_V) \; .
\label{rhoV}
\end{equation}
Here and below in Eq.~(\ref{Dyson}-\ref{H12})
the products are understood in an operator sense and
integrations over internal spatial (not time) coordinates are assumed.
Eq.~(\ref{rhoV}) is exact, accounting, for instance, for the
Pauli principle. The properties of the interaction are determined by the
correlators (\ref{VV}). Eq.~(\ref{rhoV}) had been derived in Ref.\
\onlinecite{GZB}, and below we will outline this derivation.
We can add that Eq.~(\ref{rhoV}) has been independently
re-derived by Eriksen {\it et al.}\cite{EHB}
and also by von Delft\cite{Jan}, who applied Wick's theorem.

The full $2\times 2$ matrix Keldysh  Greens function
$\check G_V(t_1,t_2,\mathbf{r}_1,\mathbf{r}_2)$ in the fluctuating
potentials $\check V=V_i\delta_{ij}$ (for $i=1,2$) satisfies the equation
\begin{equation}
\big( i\check 1\partial /\partial t_1 - \check 1 H_0 +e\check V\big)\check G_V=
\check \sigma_z\delta(t_1-t_2)\delta(\mathbf{r}_1-\mathbf{r_2}) \; .
\label{Dyson}
\end{equation}
 It follows from Eq.~(\ref{Dyson}) that the matrix elements
of $\check G_V$ can be expressed by the evolution operators
$U_{i}(t_1,t_2)={\bf T}
\exp\left[-i\int_{t_1}^{t_2} dt\;\big( H_0-eV_{i}\big) \right]$ as follows: 
\begin{eqnarray}
G_{11,V}&=&-i\theta(t_1-t_2)U_1(t_1,t_2)+iU_1(t_1,t)\rho_V(t)U_1(t,t_2)
\, ,
\nonumber\\
G_{22,V}&=&-i\theta(t_2-t_1)U_2(t_1,t_2)+iU_2(t_1,t)\rho_V(t)U_2(t,t_2)
\, ,
\nonumber\\
G_{12,V}&=&iU_1(t_1,t)\rho_V(t)U_2(t,t_2) \, ,
\nonumber\\
G_{21,V}&=&-iU_2(t_1,t)[1-\rho_V(t)]U_1(t,t_2) \, .
\label{Gij}
\end{eqnarray}
On the other hand, $\check G$ satisfies the Dyson equation
\begin{equation}
\check G_V(t_1,t_2)=\check G_0(t_1,t_2)-\int_{t_i}^t dt'
\check G_0(t_1,t')e\check V(t')\check G_V(t',t_2),
\label{Dyson1}
\end{equation}
where the Keldysh matrix $\check G_0$ is defined by Eq. (\ref{Gij}) with 
$V_{i}=0$. Eq. (\ref{Gij}) is compatible with Eq. (\ref{Dyson1}) only if
\begin{equation}
\rho_V(t)=U_1(t,t_i)\big\{1-U_1(t,t_i)\rho_0 [U_1(t_i,t)
-U_2(t_i,t)]
\big\}^{-1}\rho_0 U_2(t_i,t),
\label{solution}
\end{equation}
where $\rho_0$ is an arbitrary time-independent initial density matrix.
One can verify directly that the density matrix (\ref{solution}) satisfies
the equation of motion (\ref{rhoV}). This completes the derivation of 
(\ref{rhoV}).

The solution (\ref{solution}) in principle allows one to
derive an exact expression for the conductivity.
Alternatively, we derive here the conductivity
directly from (\ref{rhoV}). In the
presence of a weak external electric potential,
$V_x(\mathbf{r})=-\mathbf{Er}$,  Eq.~(\ref{rhoV}) can
be linearized in $V_x$ to yield
\begin{equation}
i\frac{\partial\delta\rho_V}{\partial t} = H_1\delta\rho_V-\delta\rho_V H_2-
[eV_x,\rho_V],
\label{deltarho}
\end{equation}
where $\delta\rho_V$ is proportional to $V_x$ and
\begin{eqnarray}
H_1=H_0-eV^+-\frac{1}{2}(1-2\rho_V)eV^-,\;
H_2=H_0-eV^++\frac{1}{2}eV^-(1-2\rho_V) \, .
\label{H12}
\end{eqnarray}
Continuing along the lines outlined in Section 2,
we find the conductivity,
\begin{eqnarray}
\sigma &=& \frac{e^2}{3m}
\int\limits_{-\infty}^{t_{\rm f}}dt'\int d\mathbf{r}'_{1}d\mathbf{r}'_{2} \,
[\nabla_{r_{1 {\rm f}}}-\nabla_{r_{2 {\rm f}}}]
\nonumber\\
&&
\times \bigg\langle
J_V(t_{\rm f},t',\mathbf{r}_{1 {\rm f}},\mathbf{r}_{2 {\rm f}},\mathbf{r}'_{1},\mathbf{r}'_{2}) \,
[\mathbf{r}'_{1}-\mathbf{r}'_{2}] \,
\rho_V(t',\mathbf{r}'_{1},\mathbf{r}'_{2})\bigg\rangle_{V^\pm}
\bigg|_{\mathbf{r}_{1 {\rm f}}=\mathbf{r}_{2 {\rm f}}} \, ,
\label{sigmaint}
\end{eqnarray}
where
\begin{eqnarray}
J_V(t,t',\mathbf{r}_{1 {\rm f}},\mathbf{r}_{2 {\rm f}},\mathbf{r}'_{1},\mathbf{r}'_{2})=
\int{\cal D}\mathbf{p}_1{\cal D}\mathbf{r}_1{\cal D}\mathbf{p}_2{\cal D}\mathbf{r}_2\,
{\rm e}^{iS_1[\mathbf{p}_1,\mathbf{r}_1,V^\pm]-iS_2[\mathbf{p}_2,\mathbf{r}_2,V^\pm]},
\label{JV}
\end{eqnarray}
and
\begin{eqnarray}
S_{1,2}=\int\limits_{t'}^{t_{\rm f}}dt
\bigg[\mathbf{p}\dot\mathbf{r}-\frac{\mathbf{p}^2}{2m}-U(\mathbf{r})
+eV^+(t,\mathbf{r})
\pm\frac{[1-2\rho_V(\mathbf{p},\mathbf{r})]}{2}
eV^-(t,\mathbf{r})\bigg].
\label{S12}
\end{eqnarray}
The only approximation used so far is RPA, which allows a
restriction to Gaussian fluctuations of the potentials $V^\pm$. If
one  averages over $V^\pm$ with the aid of full
nonlinear action of the electro-magnetic field, then  
Eqs. (\ref{sigmaint}-\ref{S12}) are exact. It is also useful to note
at this point that the form of the action (\ref{S12})
is not unique since the Hamiltonians (\ref{H12}) are
sensitive to the ordering of momentum and coordinate operators. 
Under these circumstances  details of  
quantization procedure become important. 
However, all these uncertainties
are of quantum origin and disappear in the quasiclassical limit $\hbar\to 0,$
which will be the only interesting limit in what follows.

 From here on we resort to further approximations.
{\sl First}, in Eq.\ (\ref{sigmaint}) we replace $\rho_V$ by the
equilibrium density matrix of non-interacting electrons $\rho_0$.
As has been pointed out in Ref.\ \onlinecite{Jan} this is
equivalent to the assumption of a
factorized initial density matrix in the CL and similar models,
an approximation which is sufficient in many cases,
although for certain correlators it is known to fail in the long-time
limit\cite{Weiss}. 
{\sl Second}, also in Eq.\ (\ref{S12}) we replace $\rho_V$ by $\rho_0$.
It is possible to show that for the problem of weak localization 
both these approximations are justified.
We will return to this point in the discussion in Section 4.
Here we only like to note that this statement
can be justified by rewriting the exact solution 
(\ref{solution}) in the form of a power series, each term of which
can be handled without replacing $\rho_V$ by $\rho_0$.

Having made these
approximations, we perform the averaging over $V^\pm$ in Eq.
(\ref{sigma3}) and get the
conductivity:
\begin{eqnarray}
\sigma = \frac{e^2}{3m}\int\limits_{-\infty}^{t_{\rm f}}dt'
\int d\mathbf{r}'_{1}d\mathbf{r}'_{2}
[\nabla_{\mathbf{r}_{1 {\rm f}}}-\nabla_{\mathbf{r}_{2 {\rm f}}}]\,
J(t_{\rm f},t',\mathbf{r}_{1 {\rm f}},\mathbf{r}_{2 {\rm f}},\mathbf{r}'_{1},\mathbf{r}'_{2})
\bigg|_{\mathbf{r}_{1 {\rm f}}=\mathbf{r}_{2 {\rm f}}}
\nonumber\\
\times\,
[\mathbf{r}'_{1}-\mathbf{r}'_{2}]\,
\rho_0(\mathbf{r}'_{1},\mathbf{r}'_{2}) \; ,
\label{sigma}
\end{eqnarray}
where
\begin{eqnarray}
J(t_{\rm f},t',\mathbf{r}_{1 {\rm f}},\mathbf{r}_{2
\rm{f}},\mathbf{r}'_{1},\mathbf{r}'_{2}) &=&
\int{\cal D}\mathbf{p}_1{\cal D}\mathbf{r}_1
\int {\cal D}\mathbf{p}_2{\cal D}\mathbf{r}_2\,
\nonumber\\
&\times&
{\rm e}^{i\tilde S_0[\mathbf{p}_1,\mathbf{r}_1]
-i\tilde S_0[\mathbf{p}_2,\mathbf{r}_2]
-iS_{\rm R}[\mathbf{p}_1,\mathbf{r}_1,\mathbf{p}_2,\mathbf{r}_2]
-S_{\rm I}[\mathbf{r}_1,\mathbf{r}_2]} \, , \,
\label{J}
\end{eqnarray}
and
$
\tilde S_0[\mathbf{p},\mathbf{r}]=\int_{t'}^{t_{\rm f}}dt
[\mathbf{p}\dot\mathbf{r}-{\mathbf{p}^2}/{2m}-U(\mathbf{r})].
$
The parts of the action $iS_{\rm R}$ and $S_{\rm I}$ describe the effect of
the bath of all electrons on the motion of a single electron. We
do not give explicit expressions for $S_{\rm R}$,  since it
vanishes for time-reversed diffusive paths
(see Ref.~\onlinecite{GZB} for details) and thus is 
irrelevant for dephasing\cite{FN}.
Here we would only note that this part of the action
contains the combination
$1-2\rho_0(\mathbf{p},\mathbf{r})=
\tanh[(H_0(\mathbf{p},\mathbf{r})-\mu)/2T],$ which is
responsible for the Pauli exclusion principle and
generates the term with $\tanh[(\xi-\omega)/2T]$ in first
order perturbation theory (\ref{AAG}).
The imaginary part of the action $S_{\rm I}$ is given
by  Eq. (\ref{SI}). It is positive
for all possible pairs of paths $\mathbf{r}_1$ and $\mathbf{r}_2$,
except for two identical ones when it is zero.
Only $S_{\rm I}$ contributes to dephasing.
In the previous section we have shown that it leads to the decay
of correlations $\propto\exp(-f_d(t)),$ where $f_d(t)$ is given by Eq.
(\ref{fd}).

The derivation of Eq.\
(\ref{sigma}) does not involve disorder averaging,
except that we have used a translationally invariant form of the
voltage correlators (\ref{IR}), which is appropriate
after averaging. On the other hand, one can demonstrate that the result
(\ref{sigma}) remains valid in the general case when
the correlators depend on two arguments separately.
 We also like to point out
that the formula (\ref{sigma}) is very similar to the evolution
equation for the density matrix of a particle interacting with
the bath of oscillators (CL).

\section{DISCUSSION}

The derivation presented above leads to the conclusion that the
dephasing time of electrons in disordered metals, as measured in
weak localization experiments, remains finite at low temperatures.
This result deviates from the earlier theoretical understanding
and accordingly has been heavily criticized. In this section
we summarize several of the arguments presented against this
conclusion and our counterarguments. This discussion should further
clarify under which circumstances the present result applies.\\

\noindent
{\bf First-order perturbation theory suggests vanishing
$T=0$ dephasing.}

Based on a first-order perturbation expansion AAG\cite{AAG}
concluded that the dephasing rate is given by
expression (\ref{AAG}). It contains the combination
$\coth(\omega/2T)+\tanh[(\xi-\omega)/2T]$, which 
vanishes in the limit $\xi,T\to 0$. 

In contrast, the 
result (\ref{GZ}) contains only $\coth(\omega/2T)$ and remains finite
for $T \rightarrow 0$. The term $\tanh[(\xi-\omega)/2T]$
in (\ref{AAG}) comes from the real part of the action
$S_{\rm R}$, which vanishes for time-reversed pairs of classical paths, and
thus is of no importance for the decay of coherence
in the long time limit. 

Perturbation theory probes only short times. In general it cannot be
used for the calculation of 
$\tau_\varphi$, since this would require an extrapolation
to long times, e.g., by exponentiation of
the first order perturbative result. Such a step is ambiguous, as
illustrated by the well-studied problem of the quantum 
decay of a metastable state in the presence of dissipation. 
The semi-classical decay rate is $\Gamma =B\exp (-A)$.
The exponent $A=A_0+\eta A_1$ is the action on a saddle-point
path\cite{CL}.  Here  the
friction coefficient $\eta$ describes the interaction  with
a dissipative bath, and $A_{0,1}$ (as well as $B_{0,1}$ below) are 
positive and independent of $\eta$.
 Fluctuations around the saddle-point path
contribute to the pre-exponent $B$. In the weak damping limit
one gets\cite{Weiss} $B\simeq B_0+\eta B_1$. It is obvious that this
result and the specific separation into prefactor and exponent -- although
it can be expanded in $\eta$ -- cannot be derived 
unambiguously from a purely perturbative calculation. 
Similarly, for the calculation of the dephasing rate it is important
to separate the pre-exponent, which is of little importance, 
from the contributions to the exponent. In contrast to the first order
calculation the saddle-point analysis
of GZ\cite{GZL,GZB} provides a definite prescription
for this separation. 

When comparing saddle-point  and  first-order
calculation one should note that the former concentrates on
contributions from time-reversed pairs of classical paths whereas the
latter effectively sums up contributions from all paths.
Therefore, after the saddle-point approximation is performed 
one cannot recover any more certain details of the first-order calculation.

We, furthermore, would like to point out that even in the framework of
perturbation theory, after exponentiation one would find a 
non-vanishing - though small - dephasing rate at $T=0$. It is not
contained in Eq.\ (\ref{AAG}) which is based on  Fermi's golden rule.
However, already first order perturbation theory in the interaction
yields additional non-golden-rule terms\cite{AAG,GZB2}.  These terms
arise due to the non-Markovian nature of the interaction, as a result of which
the usual Golden-rule replacement of $[1-\cos(\omega t)]/\omega^2$ by
a $\delta$-function fails at low temperature. The extra contributions are
usually small compared to the Golden-rule terms, but they are the leading
ones when the combination of $\coth$ and $\tanh$ terms cancel. 
These extra terms also contribute to dephasing at $T=0$.
\\

\noindent
{\bf In the ground state there is no electron scattering.}

This argument is based on the known properties of inelastic
scattering. As long as the energy of an incident particle is lower
than that of the first excited state of the scatterer,
inelastic scattering is forbidden. The internal ground state
wave function of the scatterer manifests itself only through a form- or
Debye-Waller factor. One may conclude then that
at low temperatures all electrons are in the
ground state, therefore electron-electron scattering is suppressed
and no dephasing occurs.

This argument is not conclusive for several reasons. First,
if one would describe electron-electron interaction in terms
of inelastic scattering one should take into account
that the scatterer is a macroscopic interacting many particle system
with a very small spacing between the energy eigenstates.
The temperatures where the experiments have been performed are much
higher than this level spacing.

Second, since the Coulomb interaction is long-range
each electron interacts with many other ones. A perturbative
electron-electron scattering picture is actually not appropriate.  A
more appropriate description
is in terms of collective electron degrees of freedom, for example
plasmons. Also the Nyquist noise description used here accounts for
collective properties. Within this picture all electrons form the
environment for a given electron. \\

\noindent
{\bf Quasiparticles are fully coherent.}

First we recall that the existence of exact quasiparticles is not proven,
especially in disordered metal.
Even if well-defined quasiparticles do exist, it would not rule out
dephasing of real electrons. It is the electric current of real
electrons which is measured.
An electron can be understood as a combination of an {\it infinite}
number of coherent quasiparticles (provided the latter can be constructed). 
Therefore it may be only partially coherent even at $T=0$.
A similar scenario can be analyzed rigorously in a
Caldeira-Leggett-type model, where one (singled-out) oscillator is coupled
to an infinite bath of oscillators. The single oscillator looses
phase coherence even though the eigenmodes of the total system are fully
coherent. \\

\noindent
{\bf During disorder averaging Hikami boxes
were missed.}

AAG\cite{AAG} argued that careful
disorder averaging within perturbation theory introduces
certain contributions, the `Hikami boxes'.
According to them, Hikami boxes correspond to some paths in our
approach, which we have missed. This would be responsible for the
finite dephasing at $T=0$.

On one hand, we understand that Hikami boxes do not correspond to any
classical
path. They appear as a formal feature of the perturbative many-body approach
used by AAG. There Hikami boxes are needed to restore the causality principle.
In the path integral approach causality is always maintained,
and Hikami boxes, together with unusual paths, do not appear
at all. Moreover, even within the approach of AAG Hikami boxes appear 
only in the terms generated by the action $S_{\rm R}$, which are 
irrelevant for dephasing. 

On the other hand, within our approach the dephasing rate is found to
be practically insensitive to the particular averaging procedure 
employed. For any given impurity configuration we find that
for {\it all} time-reversed paths the term $S_{\rm I}(t)$
is positive and grows with time. Thus for {\it any} pair of such paths
quantum interference is suppressed by a factor $\sim \exp (-S_{\rm I}(t))$.
Different procedures of averaging over these paths or the impurity
configurations, e.g.\ $\langle W(t) {\rm e}^{-S_{\rm I}(t)}\rangle$
or $W(t){\rm e}^{-\langle S_{\rm I}(t)\rangle}$, should yield
essentially the same value of dephasing rate, possibly differing by
prefactors of order one.\\

\noindent
{\bf The replacement $\rho_V\to \rho_0$
is uncontrolled.}

In Ref.~\onlinecite{Jan} it was argued that the replacement
$\rho_V\to \rho_0$ in Eq.~(\ref{sigmaint}) is dangerous. This replacement
can be interpreted as preparation of an artificial initial state of
the system. It might lead to spurious relaxation processes, which
could be interpreted as dephasing at $T=0.$

When analyzing this question it is important to note that we do not
study the time 
evolution of the density matrix of an electron, rather the
conductivity of the electron system.
Both evolve differently as can be seen from the comparison of
Eqs.~(\ref{rho0}) and (\ref{sigma2}). The initial density matrix
$\rho_0(\mathbf{r}_{1 {\rm i}},\mathbf{r}_{2 {\rm i}})$ in Eq.\
(\ref{rho0}) is replaced by the combination
$(\mathbf{r}'_{1}-\mathbf{r}'_{2})
\rho_0(\mathbf{r}'_{1},\mathbf{r}'_{2})$ in Eq.\ (\ref{sigma2}).
The latter can be interpreted as an effective density matrix,
which is strongly non-equilibrium
due to the presence of the factor $(\mathbf{r}'_{1}-\mathbf{r}'_{2}).$
This `non-equilibrium' form is the main reason of relaxation and dephasing,
rather than the assumption of factorized initial conditions. More generally,
one observes that in dissipative systems most of the correlators of
the type $\langle\widehat A(t)\widehat A(0)-\widehat A(0)^2\rangle$
decay with time even at $T=0$, which does not contradict to the fact that 
the equilibrium density matrix does not relax.

In spite of these arguments, one might still worry that
the replacement $\rho_V\to \rho_0$ could lead to an incorrect long-time
asymptotic and wrong conclusions about the magnitude of
$\tau_\varphi$. This indeed happens for certain correlators in exactly
solvable models\cite{Weiss}. We find, however, that it is not the
case for the conductivity in the weak localization problem.  We have already
shown in Sec.~2 that $\tau_\varphi$ is not sensitive to the replacement
if we can ignore the field $V^-$. The latter can be ignored since it is 
responsible only for the appearence of the part $S_{\rm R}$ in the action,
which gives zero contribution to $1/\tau_\varphi.$ 
We can further add that even if
we include $V^-$, the replacement $\rho_V\to \rho_0$ is justified.
The reason is that the integration in Eq.~(\ref{sigmaint}) is confined
to the region  $|\mathbf{r}'_{1}-\mathbf{r}'_{2}|<l$. In this range
the effect of interactions on the density matrix is
negligible because $l$ is  much shorter than the dephasing
length $L_\varphi$. Finally, we have developed an alternative 
approach based on a rewriting of the exact solution
(\ref{solution}) in the form of a series. Within this approach
the replacements $\rho_V\to \rho_0$ are not made, however the
final result for $\tau_\varphi$ remains unchanged.\\

\noindent
{\bf High frequency modes can be ignored at $T=0$.}

The difference between the present and earlier results for the
dephasing rate lies to a large extend in the inclusion or truncation of
high frequency modes. Their relative role can be demonstrated
if we consider a model problem of a particle interacting with a
bath of harmonic oscillators. An example of such a system is a
tunnel junction with high resistance in a circuit involving a
dissipative environment. The properties of such a system are described by the
so-called $P(E)$-theory\cite{Ingold}. The function $P(E)$ is the
Fourier transform of
the self-correlation function of the operator $\exp(i\hat\varphi),$ with
$\varphi=\int_{t_0}^t dt' eV(t')$ being the phase of the junction:
\begin{equation}
P(E)=\frac{1}{2\pi}\int dt\; {\rm e}^{J(t)+iEt},\;\;
{\rm e}^{J(t)}=
\langle {\rm e}^{i\hat\varphi(t)}{\rm e}^{-i\hat\varphi(0)}\rangle.
\label{correlator}
\end{equation}
The function $J(t)$ can be found exactly, $J(t)=-F(t)-iK(t),$ where
\begin{eqnarray}
F(t)&=&\frac{e^2}{\pi}\int_0^\infty d\omega\, {\rm
Re}[Z(\omega)]\omega \coth\frac{\omega}{2T}\,\frac{1-\cos\omega
t}{\omega^2}, \label{FF}
\\
K(t)&=&\frac{e^2}{\pi}\int_0^\infty d\omega\, {\rm Re}[Z(\omega)]
\,\frac{\sin\omega t}{\omega}.
\label{J1}
\end{eqnarray}
Here $Z(\omega)$ is the total impedance of the system tunnel
junction plus environment. Note that the function $F(t)$
is similar, although not identical to the function $f_d(t)$
introduced in (\ref{fd}). It is also possible to show
that $F(t)$ comes from the imaginary part of the action of the
environment $S_{\rm I},$ while $K(t)$ corresponds to the real part $S_{\rm R}.$

We can rewrite the function $F(t)$ approximately as a sum of two terms,
$$
F(t) \approx 2T\frac{e^2}{\pi}\int_0^T d\omega\, {\rm Re}[Z(\omega)]
\,\frac{1-\cos\omega t}{\omega^2} +
\frac{e^2}{\pi}\int_0^\infty d\omega\, {\rm Re}[Z(\omega)]
\,\frac{1-\cos\omega t}{\omega}.
$$
AAK and other
authors considered only the first, `thermal' term in the theory of weak
localization. The second term is `quantum'. It contains the
contribution of high frequencies.

We first consider the properties of the function
$F(t)$ at  $T=0$, which depend strongly on the impedance $Z(\omega).$
If ${\rm Re}[Z(\omega)]=0$ below certain frequency
$\omega_{\min},$ then $\exp[-F(t)]$ first decays with some
characteristic time $\tau_\varphi$, approaches a constant at
time $t\sim 1/\omega_{\min}$, and beyond this time
the correlations do not decay any
more. Now the role of different frequencies on the process
of dephasing becomes clear. The high frequency properties of the
impedance determine the initial stage of decay and introduce
the characteristic decay time $\tau_\varphi,$ while the low
frequency modes determine the long-time behavior of the
correlations. If the low frequency modes are absent, then the
decay of correlations stops at time $t\sim 1/\omega_{\min}.$

Finite temperature do not change this picture qualitatively. The
decay becomes stronger than at $T=0.$ At
$t>1/\omega_{\min}$ the decay of correlations stops again, but the
asymptotic value of the correlation function $\langle {\rm
e}^{\hat\varphi(t)}{\rm e}^{-i\hat\varphi(0)}\rangle$ is
smaller than for $T=0.$

On the other hand, if the function ${\rm Re}[Z(\omega)]$ is non-zero
at small $\omega,$
the correlations decay completely in the long time limit at any temperature
including $T=0.$\\

\noindent
{\bf Further arguments.}

Since the publication of Refs. \onlinecite{GZL,GZB} many more
arguments have been put forward challenging the conclusions reached in
these articles. To list a few of them:\\
-- The Pauli principle is not taken into account properly.\\
-- The electric field is not quantized.\\
-- The effect of interaction on classical paths is ignored.\\
-- Detailed balance is violated.\\
-- The term $iS_{\rm R}$ cancels $S_{\rm I}$ for $T \to 0$.\\
Here we only want to stress that
we have checked these as well as other arguments and found in all
cases that they did not apply or are in error.

\section{EXPERIMENT}

In our earlier work we have already demonstrated that our theory 
fits the data of Ref.\ \onlinecite{Webb}.
Here we want to add a comparison with some more recent experiments.
Natelson {\it et al.}\cite{Nat} found that their data for
$\tau_{\varphi}$ are not consistent with the geometry dependence
predicted by the standard theory\cite{AAK}, and for wide samples they
observed saturation of 
$\tau_{\varphi}$ at low temperatures. In the work of the Saclay-Michigan
collaboration no saturation was detected in silver\cite{SM1} and pure
gold\cite{SM2} samples down to $\approx 50$ mK, but a clear
saturation was observed in copper at temperatures one order of magnitude
higher. We will show that these results can be explained within
the framework of quantum dephasing due to electron-electron interactions.

First we note that Eqs.\ (\ref{tauphi}) are valid only in strictly
1D, 2D or 3D geometries. In practice, however,
the diffusion depends on the geometry of a particular sample
and intermediate regimes may occur.
For instance in the experiments of Refs. \onlinecite{Nat,SM1,SM2}
the samples are quasi-$1d$ (quasi-$2d$) in a standard
sense, i.e.\  both $L_{\varphi }=\sqrt{D\tau_{\varphi}}$ and
$L_{T}=\sqrt{D/T}$ exceed  the thickness $t$ and  width $w$ of the
sample.
However, the elastic mean free path $l$ is smaller
than $t$ and $w$.
Therefore the diffusion is three-dimensional at short times.
Since the main contribution to the dephasing at $T=0$
comes from high frequencies, the $3d$ expression 
\begin{equation}
\frac{1}{\tau_{\varphi 0}^{(3d)}}= \frac{e^2}{3\pi^2 \sigma_{\rm D} \,\sqrt{2D}} \left(
\frac{b}{\tau_{\rm e}}\right)^{3/2} \, 
\label{3d}
\end{equation}
applies. (The cut-off dependent numerical factor $b$ is of order 1.)
On the other hand, the temperature-dependent part of the rate, see
Eq.\ (\ref{sum}),  comes from
low frequencies or long times, when the diffusion is 
one-dimensional, and the corresponding form should be used.

In Ref.~\onlinecite{Nat} the maximum dephasing times for ten
(6 quasi-$1d$ and 4 quasi-$2d$) samples fabricated from the same material
($AuPd$) with practically the same resistivity were found to be nearly
universal $\tau_{\varphi}^{max} \approx
(0.8 \dots 2)\times 10^{-11}$ sec,  independent of the sample geometry.
In the Saclay-Michigan experiments\cite{SM1,SM2} the maximum dephasing
times were found to be 2 to 3 orders of magnitude longer than those in
Ref.\ \onlinecite{Nat}. Furthermore, $\tau_{\varphi}^{max}$ for the silver
sample was about an order of magnitude higher than that for the copper
sample in spite of similar (though not identical) parameters\cite{SM1}.
Observing this difference the authors\cite{SM1} suggested that the low
temperature saturation of $\tau_{\varphi}$ in disordered metal wires
is material dependent and not universal.

For our comparison we will use the data for 8 samples of Ref.\
\onlinecite{Nat} with nominally
identical resistivity $\rho \approx 24$ $\mu \Omega$ cm and $D
\approx 1.5 \times 10^{-3}$ m$^2$/sec (samples
C to F and H to K), 2 samples  of Ref.\ \onlinecite{SM1}
($Ag$ and $Cu$) and one gold sample ($Au$) of Ref.\ \onlinecite{SM2}
(denoted there by $AuMSU$).
Since for all samples of Ref.\
\onlinecite{Nat} the maximum dephasing times were
nearly the same, we only quote an average $\tau_{\varphi}^{max}$.
Experimental values of $\tau_{\varphi}^{max}$ and our
theoretical predictions for $\tau_{\varphi 0}$ (Eq.~(\ref{3d}) with
$b=1$) are summarized in the following Table:

\begin{center}
\begin{tabular}{|c|c|c|c|c|}
\hline
sample & $\tau_{\varphi}^{max}$ (sec)& $\tau_{\varphi 0}$ (sec)\\
\hline   C to F and H to K (averaged)\cite{Nat}&  1.3$\times 10^{-11}$  &
0.3$\times10^{-11}$\\ \hline   $Ag$\cite{SM1} & 10$\times 10^{-9}$  &
2$\times 10^{-9}$   \\ \hline   $Cu$\cite{SM1}  & 1.8$\times 10^{-9}$  &
0.3$\times 10^{-9}$\\ \hline
$Au$\cite{SM2}  & 8$\times 10^{-9}$  & 4$\times 10^{-9}$\\ \hline
\end{tabular}

\end{center}
We note that for  all the above samples the expression
(\ref{3d}) with $b=1$ gives correct order-of-magnitude estimates for
$\tau_{\varphi}^{max}$, even though these times differ by 2 to 3 orders of
magnitude.
By adjusting the cut-off parameter in the range $b \approx 0.3 \dots 0.6$
we could produce a quantitative agreement with all the mentioned
experimental values. 
Alternatively we can eliminate the ambiguity related to the cutoff
$b$ by considering the ratios between the maximum dephasing times for
different samples.  One finds:

\begin{center}
\begin{tabular}{|c|c|c|c|c|}
\hline
 & experiment & theory (eq. (\ref{3d}))\\ \hline
$\tau_{\varphi 0}^{AuPd}/\tau_{\varphi 0}^{Ag}$    & $1.3\times
10^{-3}$   & $1.5\times 10^{-3}$\\ \hline
$\tau_{\varphi 0}^{AuPd}/\tau_{\varphi 0}^{Cu}$    & $0.7\times
10^{-2}$   & $1\times 10^{-2}$\\ \hline
$\tau_{\varphi 0}^{AuPd}/\tau_{\varphi 0}^{Au}$    & $1.6\times
10^{-3}$   & $0.8\times 10^{-3}$\\ \hline
\end{tabular}
\end{center}

This agreement
strongly supports the conclusion about the universality  of the low
temperature saturation of $\tau_{\varphi}$ in disordered conductors.
To complete this summary we mention that
also our $1d$ and $2d$ expressions
for the  maximum dephasing time (\ref{tauphi}) give correct
order-of-magnitude estimates for $\tau_{\varphi}^{max}$ for the above
samples because the mean free path is comparable to the width and
thickness of the wires.

The comparison with the standard theory\cite{AAK} is much worse.
The authors of Ref.\ \onlinecite{Nat} found that the value of
$\tau_{\varphi}$ which follows from AAK for their $2d$
samples differs already at 1K significantly from their experimental
values. At the lowest temperature, $T \approx 80$ mK, this would imply
a deviation by $3$ to $4$ orders of magnitude. Only for
the thinnest samples the observed temperature dependence
was consistent with  $\tau_{\varphi} \propto T^{-2/3}$. However, in
these cases the values of
the dephasing time were such that $T\tau_{\varphi} <1$ for the whole
temperature range, whereas for the standard theory\cite{AAK} 
$T \tau_{\varphi} \gg 1$ is required.

We now analyze the temperature dependence of the dephasing
time $\tau_{\varphi}$ detected
in the experiments of Refs.\ \onlinecite{Nat} and
\onlinecite{SM1,SM2}. For quasi-$1d$ samples
Eq.\ (\ref{sum}) can be written in the following dimensionless form\cite{GZL}
\begin{equation}
\frac{\tau_{\varphi 0}}{\tau_{\varphi}}=1 +
\frac{T}{T_0}\sqrt{\frac{\tau_{\varphi}}{\tau_{\varphi 0}}}\;, \;
{\rm where}\;
T_0=\frac{R_{\rm K}\sigma_{\rm D}tw}{4
\tau_{\varphi 0}^{3/2} \sqrt{2D}}.
\label{besr}
\end{equation}
Here $R_{\rm K}= h/e^2 \simeq 25.8$ k$\Omega$ is the resistance quantum.
\begin{figure}[!h]
\centerline{\includegraphics[width=10cm]{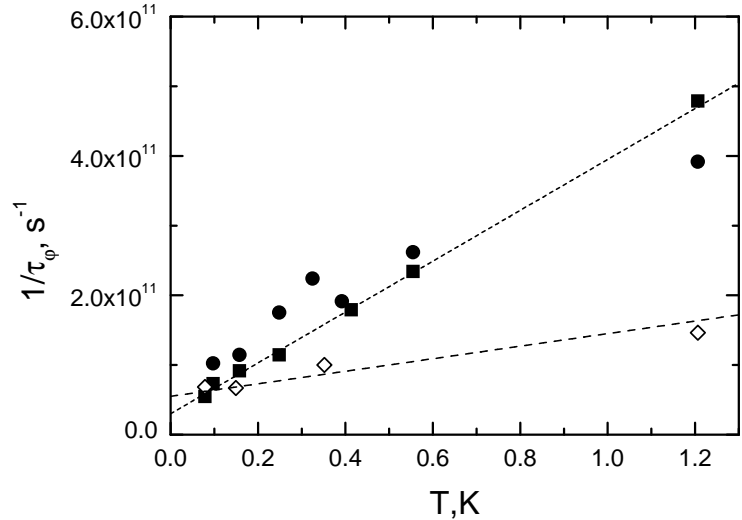}}
\caption{Dephasing rate $1/\tau_{\varphi}$ measured in
Ref. \onlinecite{Nat} for the
samples A (squares), B (circles) and F (open symbols) as a function of
temperature $T$. The straight lines are guides to the eye.}
\label{fig1}
\end{figure}
For a comparison with the power-law temperature dependence
 predicted by AAK the experimental data had been presented on a
double-logarithmic plot. However, since the temperature 
$T_0\sim 10$K is high and the dependence of Eq.\ (\ref{besr}) 
is roughly linear, 
it is more useful to re-plot them on a linear scale.
For samples with similar parameters
$\rho$, $D$ and $\tau_{\varphi 0}$ the slope
for quasi-$1d$ samples should depend only on the cross section $tw$. In
particular, since $w$ for the sample F of Ref.\ \onlinecite{Nat} is
 reported to be 4 times larger than for the samples A to E,
one expects  the slope to differ by a factor 4. The experimental
data for the samples A, B and F, presented in Fig. 1, are in good
agreement with these expectations, with a ratio $4.05$ between the slopes
corresponding to the samples A and F. On the other hand, the
magnitude of the slopes turns out to be larger than
predicted by Eq.\ (\ref{3d},\ref{besr}). The difference is as large as
 a factor of order $8$ when $b=1$, reducing to a factor $3\dots 4$
when the cut-off parameter is chosen in the range $b=0.3\dots
 0.5$. This agreement is acceptable 
within the accuracy of our theory. We conclude that our theory
accounts for the main observations of Ref.\ \onlinecite{Nat}.

Now we will address the results of Refs.\ \onlinecite{SM1,SM2}. In
Fig.\ 3 of Ref.\ \onlinecite{SM1} the authors presented their data for
$Au$, $Ag$ and $Cu$ samples. These three samples had similar classical
parameters, but both the magnitude and the temperature 
dependence of $\tau_{\varphi}$ differ drastically. This observation
led the authors to conclude 
that the behavior of $\tau_{\varphi}$ and, in particular, its saturation
at low temperature may be material dependent.
Later it was demonstrated by the same
group\cite{SM2} that the unusually low values of
$\tau_{\varphi}$ in gold\cite{SM1} were most likely due to high
concentration of magnetic impurities. A pure gold sample was
fabricated and a similar behavior was
observed\cite{SM2} as had been found previously\cite{SM1} for the $Ag$ sample:
$\tau_{\varphi}$ showed no saturation on a log-log plot down to $T
\sim 50$ mK. At the same time it was confirmed\cite{SM2}
that $Cu$ samples with very high purity showed saturation
similarly to the earlier $Cu$ sample\cite{SM1}.

A material dependence of $\tau_{\varphi}$ cannot be ruled out in general.
However, we find that such a dependence is  not
needed in order to quantitatively explain the seemingly
different behavior of $Ag$ (no saturation down to $T \sim 50$ mK) and $Cu$
(clear saturation already at $T \gtrsim 700$ mK).
The `saturation' temperature $T_0,$ defined in (\ref{besr}),
sets the scale at which thermal and quantum contributions to the
dephasing rate become comparable.
It is obvious that for otherwise identical parameters
the thinner samples will show saturation at lower temperatures.
For the parameters of the
$Cu$\cite{SM1}, $Ag$\cite{SM1} and $Au$\cite{SM2} samples
and with the cutoff parameter $b=1$
one finds from Eq. (\ref{3d},\ref{besr})
 $T_0^{Cu} \approx 2.8$ K,  $T_0^{Ag} \approx 90$ mK and $T_0^{Au}\approx 40$ mK.
It is important to note  that, for example, the ratio
$T_0^{Cu}/T_0^{Ag} \sim 30$ is not sensitive to
ambiguities related to the cutoff parameter $b$.
We conclude that the temperature where quantum effects set in
can easily vary by more than an order of magnitude for samples with similar
macroscopic parameters.

\begin{figure}
\centerline{\includegraphics[width=9cm]{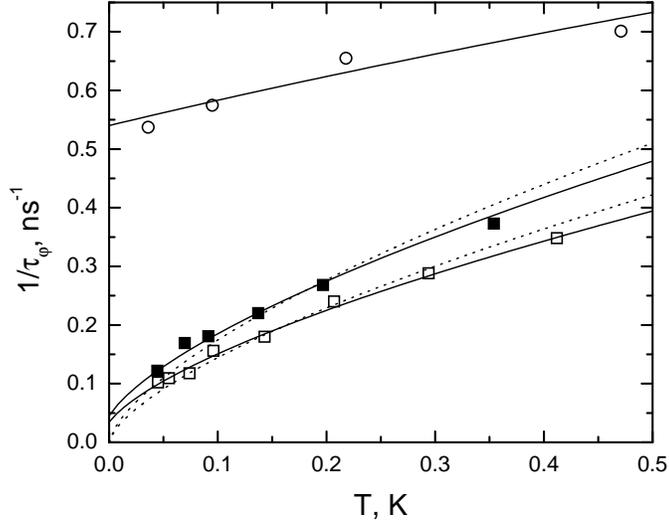}} \caption{The
values of  $\tau_{\varphi}(T)$ for the samples $Cu$
(circles), $Ag$ (open squares) of Ref. \onlinecite{SM1} and $Au$
(black squares) of Ref. \onlinecite{SM2}. Dotted lines are the best fits to the
formula\cite{AAK} $1/\tau_\varphi =AT^{2/3}.$ Solid lines are the
best fits to Eq. (\ref{besr}) with both $\tau_{\varphi 0}$
and $T_0$ varied. } \label{fig3}
\end{figure}

In Fig.\ \ref{fig3} we replot the data points for three
samples of Refs.\ \onlinecite{SM1,SM2} on a linear
scale. Best fits to the Eq. (\ref{besr}) are shown by solid
lines, while dotted lines correspond to the standard
theory\cite{AAK}. Obviously the saturating $Cu$-data 
can be fitted only with a non-vanishing dephasing rate at $T=0$.
Also the `non-saturating' samples $Ag$ and $Au$ can be fitted with a
finite $\tau_{\varphi 0}$, although one cannot draw
definite conclusions since the temperatures $T_0$ are too
low. From the fits we find $T_0^{Cu}=1.2$ K, $T_0^{Ag}=14$ mK,
$T^{Au}_0=16$ mK and $\tau_{\varphi 0}^{Cu}=2$ ns,
$\tau_{\varphi 0}^{Ag}=30$ ns, $\tau_{\varphi 0}^{Au}=20$ ns.
The deviations from the values given above can again be 
reduced if we adjust the cut-off parameter in the range 
$b=0.3 \dots 0.7.$ We conclude that the experimental data of Refs.\
\onlinecite{SM1,SM2} are consistent with the predicted low-temperature
saturation of the dephasing.
No material dependence of $\tau_{\varphi}$ needs to be assumed to get
this agreement.

\section*{ACKNOWLEDGMENT}

We thank M.~B\"uttiker, Y.~Imry, J.~Kroha, 
D.~Loss, P.~Mohanty, M.~Paalanen, J.~von Delft, 
R.A.~Webb, U.~Weiss, and P.~W\"olfle for  
stimulating discussions. 
The work is part of the {\bf CFN} (Center for
Functional Nanostructures) which is supported by the DFG (German Science
Foundation).


\begin{thebibliography}{99}

\bibitem{AGD} A.A. Abrikosov, L.P. Gor'kov, and I.Ye. Dzyaloshinkski.
{\it Quantum Field Theoretical Methods in Statistical Physics},
Second Edition, Pergamon Press 1965.

\bibitem{Bergmann} For a review see G. Bergmann, {\it Phys. Rep.} {\bf
107}, 1 (1984). 

\bibitem{AAK} B.L. Altshuler, A.G. Aronov, and D.E. Khmelnitski,
{\it J. Phys. C} {\bf 15}, 7367 (1982);
B.L.~Altshuler and A.G.~Aronov, in {\it Electron-Electron
Interactions in Disordered Systems}, eds.\ A.L. Efros and M. Pollak
(North-Holland, Amsterdam, 1985), p.1.

\bibitem{CS} S. Chakravarty and A. Schmid, {\it Phys. Rep.} {\bf 140},
193 (1986). 

\bibitem{SAI} A. Stern, Y. Aharonov, and Y. Imry, {\it Phys. Rev. A} {\bf 41},
3436 (1990). 
\bibitem{Webb} P. Mohanty, E.M.Q. Jariwala, and R.A. Webb, {\it
Phys. Rev. Lett.} {\bf 78}, 3366 (1997); {\it Fortschr. Phys.} {\bf
46}, 779 (1998). 

\bibitem{Mohanty} P. Mohanty and R.A. Webb, to be published.

\bibitem{Zawadowski} A.~Zawadowski, J.~von~Delft, and D.C.~Ralph,
{\it Phys. Rev. Lett.} {\bf 83}, 2632 (1999).

\bibitem{Nat} D. Natelson, R.L. Willett, K.W. West, and L.N. Pfeiffer,
{\it Phys. Rev. Lett.} {\bf 86}, 1821 (2001).

\bibitem{GZL} D.S. Golubev and A.D. Zaikin, {\it Phys. Rev.  Lett.}
{\bf 81}, 1074 (1998).


\bibitem{CL} A.O. Caldeira and A.J. Leggett, {\it Phys. Rev.  Lett.}
{\bf 46}, 211 (1981).

\bibitem{Weiss} U. Weiss, {\it Quantum Dissipative Systems}
World Scientific, Singapore, Second Edition (1999).

\bibitem{Loss} D. Loss and K. Mullen, {\it Phys. Rev. B} {\bf 43}, 13252 (1991).

\bibitem{GZ3} D.S. Golubev and A.D. Zaikin, {\it  Physica B} {\bf
225}, 164 (1998). 

\bibitem{Buttiker} P. Cedraschi, V. Ponomarenko, and
M. B\"uttiker, {\it Phys. Rev.  Lett.}  {\bf 84}, 346 (2000);
M. B\"uttiker, cond-mat/0106149. 


\bibitem{Peter} A.G. Aronov and P. W\"olfle, {\it Phys. Rev.  Lett.}
{\bf 72}, 2239 (1994);  {\it Phys. Rev. B} {\bf 50}, 16574 (1994).

\bibitem{GZB} D.S. Golubev and A.D. Zaikin, {\it  Phys. Rev. B} {\bf
59}, 9195 (1999). 

\bibitem{GZB2} D.S. Golubev and A.D. Zaikin,
{\it Phys. Rev. B} {\bf 62}, 14061 (2000).

\bibitem{Jan} V.~Ambegaokar and J.~von Delft, unpublished.

\bibitem{Imry} Y. Imry, {\it Introduction to Mesoscopic Physics} (Oxford
University Press, 1997).

\bibitem{AAG} I.L. Aleiner, B.L. Altshuler, and M.E. Gershenson,
{\it Waves Random Media} {\bf 9}, 201 (1999).

\bibitem{AES} U. Eckern, G. Sch\"on, and V. Ambegaokar, {\it
Phys. Rev. B} {\bf 30}, 6419 (1984).

\bibitem{EHB} K.A. Eriksen, P. Hedegaard, and H. Bruus, 
cond-mat/0101107.

\bibitem{FN} Within our quantization procedure the term $S_{\rm R}$
is purely real for all paths and, hence, cannot cancel $iS_{\rm I}$. 
Very recently von Delft\cite{vD} suggested a
slightly different procedure for which  $S_{\rm R}$ acquires also an
imaginary part. This extra part vanishes in the quasiclassical limit
$\hbar \rightarrow 0$ (see the discussion after
Eq.~(\ref{S12})). Furthermore, $S_{\rm R}$ 
averages to zero along classical
time-reversed trajectories and, hence, has no effect on $\tau_{\varphi}$. 

\bibitem{vD} J. von Delft, private communication.

\bibitem{Ingold} see e.g.\ G.-L. Ingold and Yu. V. Nazarov, in {\sl
Single Charge Tunneling}, Eds. H.\ Grabert and M.H.\ Devoret, Plenum
1992, Chapter 2.

\bibitem{SM1} A.B. Gougam, F. Pierre, H. Pothier, D. Esteve, and N.O. Birge,
{\it J. Low Temp. Phys.} {\bf 118}, 447 (2000).

\bibitem{SM2} F. Pierre, H. Pothier, D. Esteve, M.H. Devoret, A.B. Gougam,
and N.O. Birge, cond-mat/0012038.


\end{thebibliography}
\end{document}